\documentclass[10pt]{wiley2sp}
\usepackage{url}
\usepackage{helvet}
\usepackage{times}
\usepackage{amsmath}
\usepackage{graphicx}
\def\sectcounterend{.}
\def\figurename{Figure}

\def\pages{}
\def\thevol{}
\def\thenumber{}
\def\theyear{}
\def\thedoi{}

\begin{document}

\titlefigure[clip,width=.48\columnwidth]{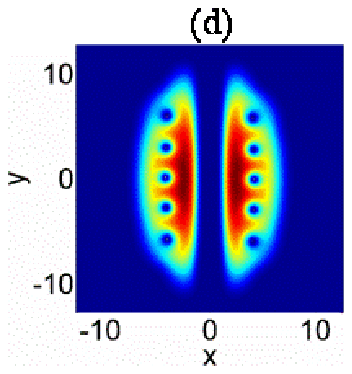}

\abstract{ We investigate the dynamics of vortex formation and the
structure of vortex lattices in a Bose-Einstein condensate confined
within a rotating double-well (DW) potential. The dynamical process
is a formation process of \textquotedblleft ghost\textquotedblright\
vortices, \textquotedblleft hidden\textquotedblright\ vortices and
\textquotedblleft visible\textquotedblright\ vortices. The critical
rotation frequency for the creation of visible vortex is indicated
by a sudden jump in the angular momentum evolution and an inflexion
in the energy evolution. Different visible vortex structures can be
formed by ruling the rotation frequency. In particular, structural
change of visible vortex patterns can be achieved by regulating the
anisotropy parameter of the DW potential. This feature allows to
flexibly control the distribution of angular momentum in macroscopic
quantum systems and study the interplay among rotation,
interparticle interaction and external potential in superfluids. }

\titlefigurecaption{Linear visible vortex lattice pair
in a Bose-Einstein condensate trapped in a rotating double-well
potential}

\title{Formation and structure of vortex lattices in a rotating double-well
Bose-Einstein condensate}

\author{L.H. Wen\inst{1*} and X.B. Luo\inst{2}}

\institute{School of Physical Sciences and Information Engineering,
Liaocheng University, Liaocheng 252059, China \and Department of
Physics, Jinggangshan University, Ji'an 343009, China}

\mail{e-mail: linghuawen@126.com}

\received{}

\keywords{Bose-Einstein condensate; vortex; double well; dynamics;
structural change}

\titlerunning{Formation and structure of vortex lattices in a rotating
DW BEC}
\authorrunning{L.H. Wen and X.B. Luo}
\pacs{03.75.Lm, 03.75.Kk, 67.85.De}
\published{}

\maketitle
\sloppy

\section{Introduction}

Topological defects have been explored extensively in many systems,
such as superfluids $^{4}$He
\cite{Donnelly} and $^{3}$He \cite{Vollhardt}, superconductors \cite%
{Blatter}, and cosmology \cite{Bhattacharjee}. Recently, increasing
interest has been focused on the topological defects in ultra-cold
atomic gases \cite{Giorgini,Fetter1,Saarikoski}. As an important
example of topological defect, quantized vortex plays a fundamental
role in the studies of atomic gases, ranging from manifesting
superfluidity \cite{Yukalov1}, collective excitations, long-range
phase coherence \cite{Yukalov2}, and universal phase transition
dynamics of Bose-Einstein condensates (BECs) and degenerate Fermi
gases \cite{Adhikari}. In particular, for the vortices in dilute
BECs, the theoretical description can be carried out in a much more
efficient way than that in liquid He due to the low density and weak
interatomic interaction.

To generate vortices in trapped BECs, different approaches, such as
stirring a BEC with a laser \cite{Madison,Abo-Shaeer}, phase
imprinting, rotating traps \cite{Hodby}, and vector gauge potential,
have been used. Here we concentrate on the case of rotating traps or
equivalently rotating BECs. It is well known that for a BEC in a
rotating harmonic trap there is a critical rotation frequency
beyond which vortex starts to exist \cite{Fetter1}. With the
increasing of rotation frequency, more and more vortices appear.
Structure of vortices depends on external trap potential. The
vortices form a triangular lattice in a symmetry harmonic trap, i.e., the celebrated Abrikosov lattice
\cite{Abo-Shaeer}. While for rotating anisotropic harmonic traps \cite%
{Oktel,Aftalion2,McEndoo1} or atomic waveguides \cite{Sinha1}, linear vortex lattices \cite{McEndoo2}, zigzag arrangements \cite%
{Gullo} and stripe-shaped configurations
\cite{Lotero,Fetter2,Matveenko2} might be generated. Furthermore,
the vortex patterns undergo structural changes depending on the
anisotropic parameter. Square vortex lattices and structural phase
transitions (from triangular lattices to square lattices) has also
been investigated in the system consisting of harmonic trap and
optical lattice \cite{Pu}. Meanwhile, the combined harmonic and
quartic traps support giant vortices \cite{Aftalion1}. The toroidal
traps can hold persistent flows with high winding numbers \cite%
{Ryu}.

To the best of our knowledge, most of the existing theories in the
literature mentioned above directly use an imaginary time
propagation (ITP) method or a lowest Landau level (LLL)
approximation or a variational approach to study the properties of a
rotating BEC. These methods are incapable of revealing the dynamics
of vortex formation although they can be applied to find the steady
states of a rotating system. In addition, these approaches can not
deal with the BECs with dissipation that is universal and inevitable
in real cold-atom experiments. Recently, a phenomenological
dissipation model has been presented to study the dynamics of vortex
formation of a BEC in a rotating harmonic trap, where elusive
\textquotedblleft ghost\textquotedblright\ vortices were revealed
\cite{Tsubota}. In a most recent investigation \cite{Wen1}, the
authors proposed an improved phenomenological dissipation model to
test the well-known Feynman rule \cite{Fetter1,Feynman} of vortices
in a rotating double-well (DW) BEC. It is found that there are three
fundamental types of vortices in a rotating DW BEC:
\textquotedblleft visible\textquotedblright\ vortex,
\textquotedblleft ghost\textquotedblright\ vortex and
\textquotedblleft hidden\textquotedblright\ vortex. Only after
including the hidden vortices can the Feynman rule be satisfied.

In this Letter, we study the dynamics of vortex formation in a BEC
trapped in a rotating DW potential by using an improved
phenomenological dissipation model \cite{Wen1}. The DW potential is
particularly important due to its simplicity, universality and yet
richness \cite{Smerzi,Wu,Luo}. Meanwhile, it is a natural and
flexible anisotropic trap. We discuss the structural change of
visible vortex patterns in a rotating DW BEC by regulating the
rotation frequency and the DW configuration. We find that the
dynamical process of vortex formation in a rotating DW potential is
remarkably different from that in a rotating harmonic trap
\cite{Tsubota}. The former case is a formation process of ghost
vortices, hidden vortices and visible vortices. The critical
rotation frequency for the generation of visible vortices is
signaled by a sudden jump in the evolution of the average angular
momentum per atom and an inflexion in the evolution of the average
energy per atom. It is shown that the rotating drive excites a
complex turbulent oscillation mainly consisting of the surface mode
with $l=4$ rather than the usual quadrupole oscillation consisting
of quadrupole mode with $l=2$ in the case of rotating harmonic trap
\cite{Tsubota}. Due to the parity effect, the vortex patterns of the
rotating DW BEC in an equilibrium state display well centrosymmetry.
By controlling the rotation frequency and the anisotropic parameter
of the DW potential, the visible vortex pattern undergoes structural
changes, including the formation of linear and zigzag vortex lattice
pairs. In particular, the linear visible vortex lattice pairs have
potential applications in quantum information processing due to the
advantages of energetic stability, long-time coherence, controllable
interaction and the least number of nearest neighbors.

\section{Phenomenological dissipation model}

By assuming strong confinement in the $z$ direction, we consider a
two-dimensional (2D) BEC trapped in a DW potential%
\begin{equation}
U(x,y)=\frac{m}{2}(\omega _{x}^{2}x^{2}+\omega
_{y}^{2}y^{2})+U_{0}e^{-x^{2}/2\delta ^{2}},  \label{DW potential}
\end{equation}%
where $m$ is the atomic mass, $\omega _{x}$ and $\omega _{y}$ are
the angular frequencies of the harmonic trap in $x$ and $y$
directions, respectively. We introduce an anisotropy parameter $\lambda =\omega _{y}/\omega _{x}$ to
describe the ratio between $\omega _{y}$ and $\omega _{x}$ . $U_{0}$
and $\delta $ denote the height and width of the potential barrier,
respectively. The DW potential is allowed to rotate around
the $z$ axis with angular frequency $\Omega $. In the rotating frame, the energy functional of the system becomes%
\begin{eqnarray}
&&E\left[ \Psi ,\Psi ^{\ast }\right]  \nonumber \\
&=&\iint dxdy\left[ \frac{\hbar ^{2}\left\vert \nabla\Psi
\right\vert ^{2}}{2m}+U(x,y)\left\vert \Psi \right\vert ^{2}-\Psi
^{\ast }(\Omega
L_{z})\Psi \right]  \nonumber \\
&&+\frac{g_{2D}}{2}\iint dxdy\left\vert \Psi \right\vert ^{4},
\label{energy functional}
\end{eqnarray}%
where $\Psi (x,y,t)$ is the wave function of the BEC, and $L_{z}=i\hbar
(y\frac{\partial }{\partial x}-x\frac{\partial }{\partial y})$ denotes the $%
z $ component of the angular-momentum operator. $g_{2D}=2\sqrt{2\pi
}\hbar ^{2}a_{s}/(a_{z}m)$ is the 2D interatomic interaction
strength with $a_{s}$ being the $s$-wave scattering length and
$a_{z}=\sqrt{\hbar /(m\omega _{z})}$ the axial harmonic length. If
the dissipation of the BEC due to thermal clouds and rotation are
ignored, one can obtain the time-dependent Gross-Pitaevskii (GP)
equation in the rotating frame by using a variational method as follow
\begin{eqnarray}
i\hbar \frac{\partial \Psi }{\partial t} &=&\left[ -\frac{\hbar ^{2}}{2m}%
\left( \nabla _{x}^{2}+\nabla _{y}^{2}\right) +U(x,y)-\Omega L_{z}\right]
\Psi  \nonumber \\
&&+g_{2D}\left\vert \Psi \right\vert ^{2}\Psi .  \label{GPE}
\end{eqnarray}%
The wave function is normalized as $N=\iint \left\vert \Psi \right\vert
^{2}dxdy$ with $N$ being atom number. The stationary
solutions $\Psi (x,y,t)=\Phi (x,y)e^{-i\mu t/\hbar }$ satisfy the time-independent
GP equation%
\begin{eqnarray}
\mu \Phi &=&\left[ -\frac{\hbar ^{2}}{2m}%
\left( \nabla _{x}^{2}+\nabla _{y}^{2}\right) +U(x,y)-\Omega
L_{z}\right]
\Phi  \nonumber \\
&&+g_{2D}\left\vert \Phi \right\vert ^{2}\Phi,
\label{time-independent GPE}
\end{eqnarray}%
where $\mu $ is chemical potential.

Equation (\ref{time-independent GPE}) or the imaginary time
propagation of the time-dependent version (\ref{GPE}) are widely
used to study the properties of the stationary states of a rotating
BEC. However, Eq. (\ref{GPE}) and Eq. (\ref{time-independent GPE})
can not reveal the dynamical process of the vortex formation. In
Ref. \cite{Feder}, the authors numerically solved the time-dependent
GP equation in a rotating frame, but the motion of created vortices
remains turbulent
and no vortex lattice forms. Moreover, Eq. (\ref{GPE}) and Eq. (\ref%
{time-independent GPE}) can not describe a BEC with dissipation.

Here we use a phenomenological dissipation model \cite{Wen1} to
study the dynamics of vortex generation and the structure of the
equilibrium state of rotating DW BEC. In this phenomenological
model, the wave function $\Psi (x,y,t)$ obeys the following modified GP equation%
\begin{eqnarray}
\left( i-\gamma \right) \hbar \frac{\partial \Psi }{\partial t} &=&\left[ -%
\frac{\hbar ^{2}}{2m}\left( \nabla _{x}^{2}+\nabla _{y}^{2}\right)
+U(x,y)-\Omega L_{z}\right] \Psi  \nonumber \\
&&+g_{2D}\left\vert \Psi \right\vert ^{2}\Psi ,  \label{dissipative GPE}
\end{eqnarray}%
where $\gamma $ describes the degree of dissipation of the BEC. This
model is a variation of that in Ref. \cite{Tsubota} and has good
predictive power. Our computation results for the case of a BEC in rotating harmonic trap are well in agreement with the experimental
observations in Ref. \cite{Madison} and the numerical results in Ref. \cite%
{Tsubota}. Recently several other phenomenological models have also been
proposed to deal with the BECs with dissipation \cite%
{Choi,Jackson,Gardiner,Hsueh}. To numerically simulate Eq. (\ref{dissipative GPE}), it is convenient to introduce
dimensionless parameters as following $d_{0}=\sqrt{\hbar /2m\omega _{x}},\widetilde{x}=x/d_{0},%
\widetilde{y}=y/d_{0},\sigma =\delta /d_{0},\widetilde{t}=t\omega
_{x},V=U/\hbar \omega _{x},V_{0}=U_{0}/\hbar \omega _{x},\widetilde{\Omega }%
=\Omega /\omega _{x},\widetilde{L_{z}}=L_{z}/\hbar $ and $\psi =\Psi d_{0}/%
\sqrt{N}$. Thus Eq. (\ref{dissipative GPE}) is reduced to a dimensionless
form as%
\begin{eqnarray}
(i-\gamma ) \frac{\partial \psi }{\partial t} &=&\left[ -\left(
\frac{\partial ^{2}}{\partial x^{2}}+\frac{\partial ^{2}}{\partial
y^{2}}\right)
+V(x,y)-\Omega L_{z}\right] \psi  \nonumber \\
&&+c\left\vert \psi \right\vert ^{2}\psi ,  \label{dimensionless
DissGPE}
\end{eqnarray}%
where $c=g_{2D}N/(\hbar \omega _{x}d_{0}^{2})$ denotes the dimensionless
nonlinear strength and\ the tilde is omitted for simplicity.\ The rescaled 2D
DW potential is expressed by%
\begin{equation}
V(x,y)=\frac{1}{4}(x^{2}+\lambda ^{2}y^{2})+V_{0}e^{-x^{2}/2\sigma ^{2}}.
\label{rescaled DW potential}
\end{equation}

We numerically solve Eq.
(\ref{dimensionless DissGPE}) with the split-step Fourier method
\cite{Wen2,Xiong}. The initial quantum state $\psi (x,y,t=0)$ in the
DW potential can be obtained by the ITP method \cite{WuB,Wen3,Zhang}
for $\Omega =0$. In our numerical computations, we use $\
^{87}$Rb atoms and the parameters are chosen as $%
\omega _{x}=\omega _{y}=2\pi \times 40$ Hz, $\omega _{z}=2\pi \times 800$
Hz, $V_{0}=40$ and $\sigma =0.5$. Here we take $\gamma =0.03$, which
corresponds to a temperature of around $0.1T_{c}$ \cite{Choi}. In fact, the
variation of the nonzero $\gamma $ in Eq. (\ref{dimensionless DissGPE}) only
influences the relaxation time scale but does not change the dynamics of the
vortex formation and the ultimate steady structure of the system.

\section{Dynamics of vortex formation}

To reveal the dynamics of vortex formation in the rotating DW BEC
for a
fixed $\Omega $, we show the time evolution of the density distribution $%
\left\vert \psi \right\vert ^{2}$ and that of the corresponding
phase distribution of $\psi (x,y,t)$ in Fig. 1 and Fig.2,
respectively. In Fig.2, the value of the phase varies continuously
from $0$ to $2\pi $, and the end point of the boundary between a
$2\pi $ phase line and a $0$ phase line corresponds to a phase
defect (i.e., a vortex). Initially, the condensate density
$\left\vert \psi (x,y,t=0)\right\vert ^{2}$ in a stationary DW
potential is displayed in Fig. 1(a), and there is a zero relative
phase between the two halves of the BEC as shown in Fig. 2(a). With
the development of time, the BEC undergoes complex turbulent
oscillation [see Fig. 1(b)] instead of the usual elliptic
oscillation in the case of
rotating harmonic trap \cite{Tsubota}. In Ref. \cite%
{Tsubota}, the elliptic oscillation is mainly caused by the
quadrupole mode with angular momentum $l=2$. Since the DW system has
even parity with respect to the spatial coordinate, no surface modes
with odd $l$ are excited. Here the turbulent oscillation principally
consists of the surface mode with $l=4$ coupling with higher-energy
modes through the nonlinear interaction, which is verified by the
critical rotation frequency for visible vortex generation. The
turbulent oscillation makes the boundary surfaces of the BEC
unstable and excites the surface waves that propagate along the
surfaces. Essentially, the surface ripples are resulted from the
dynamical instability \cite{Sinha2} associated with the imaginary
frequency of the excitation modes. Then the surface waves develop
into the vortex cores and most of the phase defects appear at the
boundary surfaces of the cloud as shown in Fig. 2(b). Since the
amplitude $\left\vert \psi \right\vert $ on the outskirts of the BEC
is almost negligible, these phase defects can not be seen in the
density distribution of Fig. 1(b) and neither carry angular momentum
nor energy, they are called ghost vortices \cite{Tsubota,Wen1}.

At the same time, there are some phase singularities congregating
and distributing symmetrically along the central barrier [see Figs.
2(b)-2(c)]. These phase singularities are known as the hidden
vortices \cite{Wen1} because they carry evident angular momentum
despite being invisible in the density distributions of Figs.
1(b)-1(c). With the further time evolution some
ghost vortices penetrate into the BEC via the Landau instability \cite%
{Dalfovo} associated with the negative excitation frequency and
becomes the usual visible vortices as shown in Figs. 1(d)-1(e) and
Figs. 2(d)-2(e). In the presence of dissipation, a steady triangular
lattice pair of visible vortices [see Fig. 1(f)] forms eventually
due to the competition between the rotating drive propelling
vortices toward the rotation axis and the repulsive interaction
tending to push the vortices apart, where the energy of the system
reaches the minimum in the rotating frame. The visible vortex
density is determined by the parameters $\Omega $, $\lambda $,
$V_{0}$, $\sigma $ and $c$. As displayed in Fig. 2(f), most of the
ghost vortices are repelled to the outside of the cloud. Since the
DW potential is even parity concerning to the spatial coordinate,
all the numbers of the visible vortices, hidden vortices and ghost
vortices should be even, which can be seen clearly in Fig. 1(f) and
Fig. 2(f). Physically, since vortex is a topological singularity,
the system undergoes the change of topological property between the
BEC with vortices and that without vortices when the external
rotating drive inputs angular momentum into the BEC. Therefore the
vortices are easier generated in the area of the lower density, such
as the barrier boundary and the outskirts of the cloud [see Figs.
1(b)-1(c) and Figs. 2(b)-2(c)]. This is the reason why the hidden
vortices and ghost vortices always appear before the visible
vortices.

\begin{figure}[tp]
\centerline{\includegraphics*[width=7.6cm]{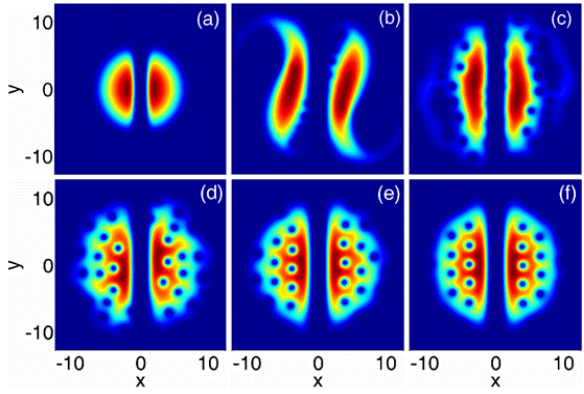}}
\caption{(online color) Time evolution of the density distribution $%
\left\vert \protect\psi \right\vert ^{2}$ after the double-well
potential with $\protect\lambda =1$ suddenly begins to rotate with
$\Omega =0.9$, where $c=600$. The time is (a) $t=0$, (b) $t=10$, (c)
$t=50$, (d) $t=100$, (e) $t=160$, and (f) $t=250$. The darker color
area indicates the lower
density. Here $x$ and $y$ are in units of $d_{0}$, and $t$ is in units of $1/%
\protect\omega _{x}$.} \label{Figure1}
\end{figure}
\begin{figure}[tp]
\centerline{\includegraphics*[width=8cm]{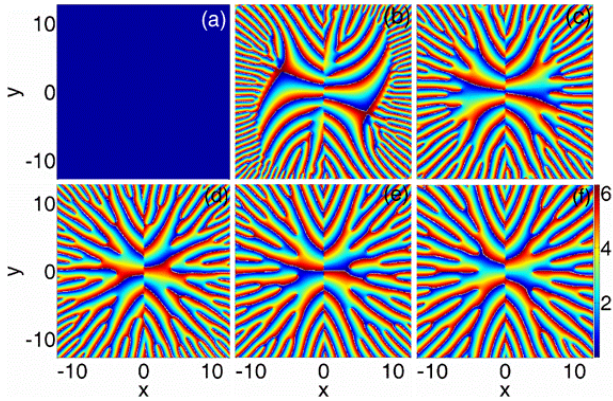}}
\caption{(online color) Time evolution of the phase distribution of
$\protect\psi (x,y,t)$ corresponding to that of the density
distribution of Fig. 1. The value of the phase varies continuously
from $0$ to $2\protect\pi $. The darker color area indicates the
lower phase. } \label{Figure2}
\end{figure}

In Figs. 3(a)-3(b), we display the time evolution of the average
angular momentum per atom $l_{z}=\iint \psi ^{\ast }L_{z}\psi dxdy$
and the average energy per atom $\varepsilon =$ $\iint \psi ^{\ast
}[-(\nabla _{x}^{2}+\nabla _{y}^{2})+V(x,y)+(c/2)\left\vert \psi
\right\vert ^{2}-\Omega L_{z}] \psi dxdy$, respectively. Here
$\lambda =1$, and $\Omega $ or $c$ have different values. When $%
\Omega <0.59$, we can see that $l_{z}$ undergoes damped oscillation
until a convergent value and $\varepsilon $ decreases monotonously
to an equilibrium value. When $\Omega \geq 0.59$, however, $l_{z}$
first experiences damped oscillation, then increases to a new peak
value, and finally descends damply to an equilibrium value [see Fig.
3(a)]. At the same time, $\varepsilon $ first decreases, then passes
over an energy barrier, and finally drops gradually to a convergent
value [see Fig. 3(b)]. This distinct difference indicates that the
critical rotation frequency for the creation of visible vortices is
$\Omega _{c}=0.59$ which characterizes the topological phase
transition of the rotating DW system. The critical frequency $\Omega
_{c}=0.59$ agrees well with that predicted in Ref. \cite{Wen1}.
However, this value is lower than the usual critical frequency
$\Omega _{hc}=\sqrt{2}/2\simeq 0.707$
predicted for a rotating harmonic trap due to the dynamical instability \cite%
{Tsubota,Stringari}. In view of the even parity of the DW potential,
the most probable reason is that here the rotating drive excites
mainly the high-order surface mode with $l=4$ rather than the
quadrupole mode with $l=2$, where $\Omega _{hc}=\sqrt{l}/l$. The
small difference between $\Omega _{c}=0.59$ and $\Omega _{hc}=0.5$
is due to the nonlinear interatomic interaction and the presence of
the Gaussian barrier as well as the resulting equivalent anisotropy.
Moreover, the larger is
the $\Omega $ (or $c$) with the same $c$ (or $\Omega $), the larger is the $%
l_{z}$ in the equilibrium state and the lower (or higher) is the
corresponding $\varepsilon $ with a slower (or faster) decline,
which reveals that the vortex array is a kind of locally stable
quantum state. As shown in Fig. 1 and Fig. 2, with the development
of time the hidden vortices and some ghost vortices penetrate toward
the areas of high density in the cloud. Due to the DW potential
consisting of a harmonic trap and a Gaussian barrier, the vortex
formation process is an oscillating dynamical process concerning to
the inputting of the external angular momentum, which can explain
the oscillation behavior of the angular momentum in Fig. 3(a).

\begin{figure}[tp]
\centerline{\includegraphics*[width=8.1cm]{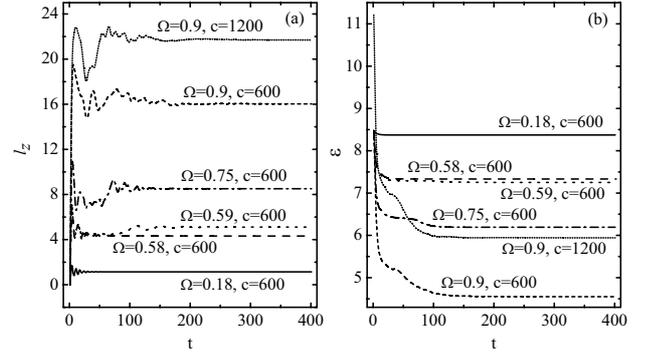}}
\caption{Time evolution of (a) the average angular momentum per atom
$l_{z}$
and (b) the average energy per atom $\protect\varepsilon $ for $\protect%
\lambda =1$. Here $t$, $l_{z}$ and $\protect\varepsilon $ are in
units of
$1/\protect\omega _{x}$, $\hbar $ and $\hbar \protect%
\omega _{x}$, respectively. } \label{Figure3}
\end{figure}

\section{Structural change of vortex patterns}

The variation of the rotation frequency influences not only the
dynamics of the vortex formation but also the equilibrium structure
especially the visible vortex pattern of the rotating DW system. In
Fig. 4, we display the steady density distributions $\left\vert \psi
\right\vert ^{2} $ at $t=250$ after the DW potential rotates with
different angular frequencies, where $\lambda =1$ and $c=600$. The
critical rotation frequency at which visible vortices can enter the
BEC with $\lambda =1$ is $\Omega _{c}=0.59$ as has been pointed out
above and in Ref. \cite{Wen1}. From Fig. 4(a), there is a visible
vortex pair for the case of $\Omega =0.65$, which is similar to the
case of $\Omega =0.59$ \cite{Wen1}. By comparison, the $l_{z}$ in
the equilibrium state for the former case is larger than that for
the latter case (see Fig. 3 in \cite{Wen1}), which is caused by the
further approach of the visible vortices to the center of the DW
potential and the increase as well as the further concentration of
the hidden vortices distributing along the central barrier (the
phase profile for $\Omega =0.65$ is not shown here). With the
increase of $\Omega $, the visible vortices enter the BEC with the
pair shape and form linear vortex lattice pair as shown in Figs.
4(b)-4(d). When $\Omega =0.85$, the mirror symmetry is broken and a
zigzag vortex lattice pair is formed [see Fig. 4(e)]. Finally, the
transition into a triangular vortex lattice pair occurs, which can
be seen clearly in Fig. 4(f) and Fig. 1(f). Here a common
characteristic is that these visible vortex patterns exhibit the
centrosymmetry due to the even parity of the DW potential with
respect to the spatial coordinate.

\begin{figure}[tp]
\centerline{\includegraphics*[width=7.6cm]{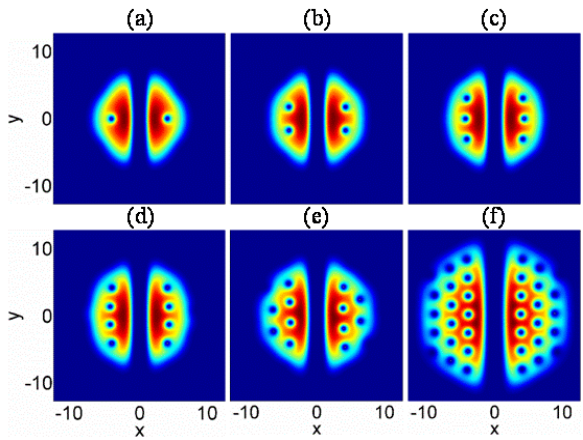}}
\caption{(online color) Steady density distributions $\left\vert \protect%
\psi \right\vert ^{2}$\ at $t=250$ for different rotation frequencies: (a) $%
\Omega =0.65$; (b) $\Omega =0.7$; (c) $\Omega =0.75$; (d) $\Omega
=0.8$; (e) $\Omega =0.85$; (f) $\Omega =0.95$. Here $\protect\lambda
=1$ and $c=600$. The darker color area indicates the lower density.}
\label{Figure4}
\end{figure}

In Fig. 5, we present the steady density distributions $\left\vert
\psi \right\vert ^{2}$ at $t=250$, where $c=600$ and $\lambda $
takes on different values. For the fixed angular frequency $\Omega
=0.9$, when $\lambda =2$ the visible vortices form a
linear lattice pair along the $x$ axis due to the tight confinement in the $%
y $-direction of the trap as shown in Fig. 5(a). When $\lambda
=1.5$, the visible vortices become a zigzag vortex lattice pair [see
Fig. 5(b)]. A further reduction of $\lambda $ leads to a formation
of a triangular lattice pair of the visible vortices [see Fig. 5(c)
and Fig. 1(f)]. The visible vortex patterns for the fixed angular frequency $\Omega =0.8$ with different values of $%
\lambda $ are illustrated in Fig. 4(d) and Figs. 5(d)-(f). With the
decrease of $\lambda $, the visible vortices entering the BEC by
means of vortex pair arrange in a linear lattice pair along the
$y$-direction because of the strong restraint in the $x$-direction
of the trap. In particular, the two halves of the BEC approximate
two rectangles at $\lambda =0.8$, where the linear visible vortex
lattice pair are similar to two ionic crystals \cite{Birkl}.

\begin{figure}[tp]
\centerline{\includegraphics*[width=7.6cm]{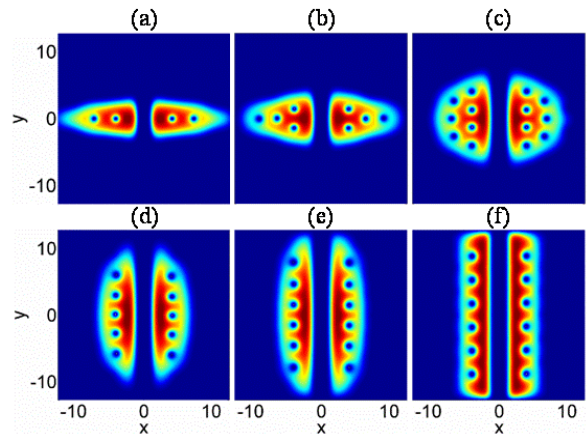}}
\caption{(online color) Steady density distributions $\left\vert \protect%
\psi \right\vert ^{2}$\ at $t=250$ for different values of $\protect\lambda $%
, where $c=600$. (a) $\protect\lambda =2,$ $\Omega =0.9$, (b) $\protect%
\lambda =1.5,$ $\Omega =0.9$, (c) $\protect\lambda =1.1,$ $\Omega
=0.9$, (d) $\protect\lambda =0.9,$ $\Omega =0.8$, (e)
$\protect\lambda =0.85,$ $\Omega =0.8$, and (f) $\protect\lambda
=0.8,$ $\Omega =0.8$. The darker color area indicates the lower
density.} \label{Figure5}
\end{figure}

From Fig. 3(f) in Ref. \cite{Wen1} and Fig. 4(a), we can see that
for $0.59$ $\leq \Omega \leq 0.65$ there are always two visible
vortices, i.e., the number of the visible vortices is degenerate
with respect to the rotation frequency. Similar degeneracy also
occurs in other certain intervals of rotation frequency. In fact, we
find that the visible vortex number is possibly degenerate
concerning to not only $\Omega $ but also $\lambda $, for instance,
there are all the twelve visible vortices for the cases of $0.8$
$\leq \lambda \leq 0.85$ with the same $\Omega =0.8$ as shown in
Figs. 5(e)-5(f). The vortex patterns in Fig. 4 and Fig. 5
are rather stable as verified in our simulation. Recently the
stability of linear vortex lattice in a
rotating anisotropic harmonic trap has been discussed in Ref. \cite%
{McEndoo1}.

The structure change of vortex patterns in a BEC in a rotating anisotropic
harmonic trap has been studied by several theoretical groups \cite%
{Oktel,Aftalion2,Sinha1,McEndoo2,Gullo,Lotero,Fetter2,Matveenko2},
where the authors adopted mainly the LLL approximation or
Thomas-Fermi (TF) approximation to analyze the vortex structures.
However, all of these works did not consider the inevitable
dissipation of rotating BECs. In
addition, the analytical approaches in Refs. \cite%
{Oktel,Aftalion2,Sinha1,McEndoo2,Gullo,Lotero,Fetter2,Matveenko2} can not be used directly to investigate the steady-state structures
of a rotating DW BEC due to the presence of the Gaussian barrier.
Here, for the rotating DW BEC there are three different kinds of
vortices: visible vortices, hidden vortices and ghost vortices, and
each of them have their own unique topological structures [see Fig.
1(f), Fig. 2(f), and Fig. 4], which is very different from the cases
of rotating anisotropic harmonic trap. Moreover, as the values of
all the parameters are provided above, the structural change of the
visible vortex patterns for the rotating DW BEC with different
$\Omega $ or different $\lambda $ is experimentally accessible and
examinable.

On the other hand, the linear visible vortex lattice pairs in Fig. 4
and Fig. 5 have potential applications in quantum information as
they are somewhat similar to the ionic crystals \cite{Birkl}. For
instance, a visible vortex in linear arrangement may be encoded as a
quantum bit via a
vortex-antivortex superposed state \cite%
{Wen2,Wen3,Kapale,Thanvanthri,Liu}. The flip between the vortex and
the antivortex can be realized either by optical methods
\cite{Andersen,Wright} or by compressing and rehabilitating the
trapping potential \cite{Liu}. The coherence time of vortex quantum
bit is
long enough to allow a series of operations \cite%
{McEndoo1}. In view of the degeneracy of visible vortex number, the
small errors of $\Omega $ or $\lambda $ due to experimental
uncertainties do not influence the manipulation toward the vortex
bits, which is beneficial to the information processing. Another
advantage of the linear visible vortex lattices is that each visible
vortex has only two nearest neighbors with controllable interaction,
where the effect caused by the distant hidden or ghost vortices can
be ignored because the particle density in the regions of the
potential barrier and the outskirts of the cloud is very small.

\section{Conclusion}

In summary, we have numerically studied the dynamics of vortex
formation and the structure of vortex lattices in a
rotating DW BEC by using a phenomenological dissipation model
\cite{Wen1}. The vortex formation process in a rotating DW BEC is
evidently different from that in a rotating single-well BEC
\cite{Tsubota}. For the latter case, there are at most two types of
vortices: visible vortex and ghost vortex. For our case, there exist
three kinds of vortices: visible vortex, ghost vortex and hidden
vortex. Once the DW potential rotates, the BEC undergoes complex
turbulent oscillation principally consisting of the surface mode
with $l=4$ instead of the usual elliptic oscillation consisting of
quadrupole mode with $l=2$ in a rotating harmonic trap
\cite{Tsubota}. The hidden vortices and ghost vortices always form
earlier than the visible vortices. The critical frequency for the
visible vortex generation is marked by a sudden jump in the
evolution of the average angular momentum per atom and an inflexion
in the evolution of the average energy per atom. The steady vortex
patterns of the rotating DW BEC display well centrosymmetry because
of the parity effect. It is shown that a structure change of the
visible vortex patterns (including the formation of linear and
zigzag vortex lattice pairs) can be realized by governing the
rotation frequency and the anisotropic parameter of the DW
potential. The structural transition of visible vortex patterns
makes it possible to artificially manipulate the distribution of
vorticity in superfluids and apply the quantized vortices to quantum
information processing.
\\

\emph{Acknowledgments} We thank Biao Wu and Hongwei Xiong for
helpful discussions, and thank Yongping Zhang for valuable
suggestions and carefully reading our manuscript. This work is
supported by the NSFC (Grants No. 11047033, No. 10965001 and No.
11165009), the International Cooperation Program by Shandong
Provincial Education Department, the NSF of Jiangxi Province under
Grant No. 2010GQW0033, and the Jiangxi Young Scientists Training
Plan under Grant No. 20112BCB23024.

\end{document}